\documentclass[11pt]{article}

\usepackage[utf8]{inputenc}
\usepackage[T1]{fontenc}
\usepackage{amsmath,amssymb}
\usepackage{graphicx}
\usepackage{booktabs}
\usepackage{hyperref}
\usepackage{listings}
\usepackage{xcolor}
\usepackage[margin=1in]{geometry}
\usepackage{tikz}
\usetikzlibrary{arrows.meta,positioning,fit,backgrounds,calc}

\title{Threadle: A Memory-Efficient Network Storage and Query Engine for Large, Multilayer, and Mixed-mode Networks}

\author{
  Carl Nordlund\textsuperscript{1,*} \and Yukun Jiao\textsuperscript{1} \\[1ex]
  \textsuperscript{1}The Institute for Analytical Sociology, Department of Management and Engineering, \\
  Linköping University, Norrköping, Sweden \\[1ex]
  \textsuperscript{*}Corresponding author: carl.nordlund@liu.se
}

\date{}

\begin{document}

\maketitle

\begin{abstract}
We present Threadle, an open-source, high-performance, and memory-efficient network storage and query engine written in C\#. Designed for working with full-population networks derived from administrative register data, which represent very large, multilayer, mixed-mode networks with millions of nodes and billions of edges, Threadle addresses a fundamental limitation of existing network libraries: the inability to efficiently handle two-mode (bipartite) data at scale. Threadle's core innovation is a pseudo-projection approach that allows two-mode layers to be queried as if they were projected into one-mode form, without ever materializing the memory-prohibitive projection. We demonstrate that a network with 20 million nodes containing layers equivalent to 8 trillion projected edges can be stored in approximately 20 GB of RAM—a compression ratio exceeding 2000:1 compared to materialized projection. Additionally, Threadle provides native support for multilayer mixed-mode networks, an integrated node attribute manager, and a CLI frontend with 50+ commands for the construction, processing, file handling, and management of very large heterogeneous networks. Threadle is freely available at \url{https://www.threadle.dev} and can either be obtained as precompiled binaries for Win, macOS and Linux, or compiled directly from source. Supplementing Threadle is threadleR, an R frontend that enables advanced sampling- and traversal-based analyses on very large, heterogeneous, multilayer, mixed-mode population-scale networks.\end{abstract}

\section{Introduction}

Large-scale administrative register data, such as national population registers used for research, can be represented as large, temporal, and structurally complex networks containing multiple relational layers. Kinship, residence, employment, and education all constitute distinct relational domains, many of which are naturally represented as two-mode (bipartite, affiliation) structures: individuals are connected not directly to each other, but through shared membership in households, workplaces, schools, and residential areas.

Research applications involving multilayer random-walker models, simulation-based methods, and repeated sampling of ego networks across layers require very fast retrieval of neighboring alters and node attributes, while maintaining a memory footprint small enough to allow population-scale networks for multiple years to be simultaneously loaded into RAM. Given these requirements, analyses on population-scale networks necessarily rely on sample- and traversal-based methods rather than exhaustive computation, making efficient storage and fast querying the critical requirements.

The fundamental computational challenge is the projection problem. Two-mode data are commonly projected to one-mode form for analysis, expanding each affiliation of $k$ nodes into $\binom{k}{2} = k(k-1)/2$ edges, i.e. node pairs. For large-scale register data, this expansion can be catastrophic: a single affiliation type (e.g., workplace co-affiliation) for an entire national working population can produce billions of projected edges, while residential co-location would produce orders of magnitude more. Materializing such projections exhausts available RAM on any practical system.

Threadle was developed to address this challenge, providing a dedicated backend for representation and querying of full-population, feature-rich networks with multiple relational layers of different properties and modes. Its core innovation—pseudo-projection—allows two-mode data to be queried as if projected without ever computing the projection, reducing memory requirements by several orders of magnitude.

\section{State of the Field}

General-purpose network libraries such as igraph \cite{igraph}, NetworkX \cite{networkx}, NetworKit \cite{staudt_networkit_2016}, and graph-tool \cite{peixoto_graph-tool_2017} offer extensive analytical toolsets, but their internal data models are not designed to support the structural and memory requirements of multilayer and mixed-mode data found in full-population register data. Specifically, we note the following limitations:

\begin{itemize}
    \item \textbf{Multilayer representation}: Networks with multiple relational types are typically represented by attaching attributes to edges rather than storing them natively within layers in the engine, incurring substantial memory overhead.
    
    \item \textbf{Unipartite assumption}: Relations are typically stored as unipartite graphs where all relations across layers and edge types must be represented as node-pairs in a single one-mode edgelist.
    
    \item \textbf{Projection requirement}: Two-mode (bipartite) data are commonly projected to their one-mode form, expanding each affiliation of $k$ nodes into $k(k-1)/2$ pairwise edges, which is memory-prohibitive for large bipartite datasets.
    
    \item \textbf{Attribute storage}: Node attributes are often stored in general-purpose metadata containers in R or Python rather than within the graph engine itself.
    
    \item \textbf{Algorithmic complexity}: Many commonly used methods and metrics, such as betweenness centrality, closeness centrality, and community detection, become computationally infeasible for very large networks due to their time and memory complexity, although approximation algorithms exist for some measures \cite{staudt_networkit_2016}.
\end{itemize}

These limitations are architectural rather than incidental: the existing libraries are designed around unipartite graph representations with analytical algorithms as their primary focus. Contributing native multilayer and mixed-mode storage to these established codebases would require fundamental restructuring of their core data models. Threadle therefore takes a complementary approach: rather than replacing these analytical frameworks, it provides an alternative storage and query layer optimized for large, feature-rich networks, where sample- and traversal-based analytical methods are implemented at the frontend level.

\section{Software Architecture}

\begin{figure}[ht]
    \centering
    \includegraphics[width=0.8\textwidth]{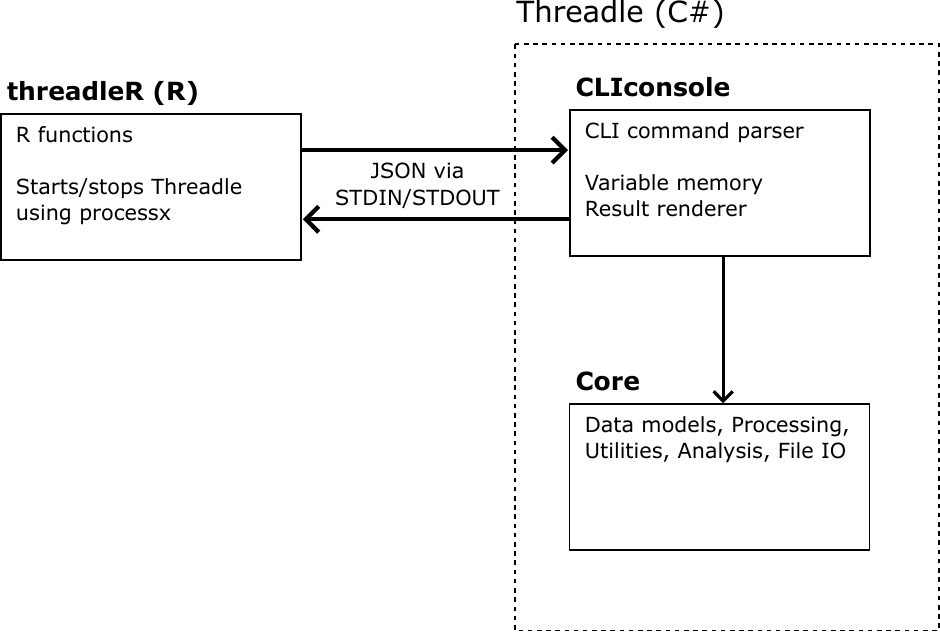}
    \caption{Threadle system architecture. Threadle.Core implements all data structures and methods as a .NET 8.0 library. Threadle.CLIconsole exposes this functionality through a scripting language with text and JSON modes. The threadleR package provides seamless R integration via JSON mode, enabling researchers to combine Threadle's efficient storage with R's statistical capabilities. Threadle.Core can also be embedded directly as a project reference in existing C\# solutions.}
    \label{fig:architecture}
\end{figure}

Threadle's architecture comprises three principal components, illustrated in Figure~\ref{fig:architecture}.

\subsection{Threadle.Core}

Threadle.Core is a .NET 8.0 library implementing all data models, storage structures, processors, methods, and file I/O. The two fundamental structure types are Nodesets and Networks, both implementing a shared interface.

A \textbf{Nodeset} is a lightweight collection of node identifiers with optional node attributes. Nodes are identified by unique unsigned integers and stored in either a hashset or dictionary structure depending on whether they have attributes.

This dual-storage design reflects the realities of administrative register data, where attribute availability is often heterogeneous: birth year exists for all individuals, but income only for adults, educational attainment only for those who completed schooling, and workplace identifiers only for the employed. Storing absent attributes as null values would waste substantial memory at population scale. Instead, Threadle's node attribute manager supports four compact types—32-bit integer, 32-bit float, boolean, and single character—and stores attribute values only for nodes that possess them. Nodes automatically migrate between the lightweight hashset (no attributes) and dictionary structure (with attributes) as attributes are added or removed, ensuring minimal memory overhead regardless of attribute sparsity.

A \textbf{Network} references a Nodeset and consists of layers of relations, where each node owns its edges or hyperedge memberships within each layer. Layers are defined as either one-mode or two-mode, with one-mode layers supporting configurable directionality, edge valuation, and self-tie allowance.

Threadle.Core is organized into four functional areas: Model (data structures for Nodesets, Networks, layers, edgesets, hyperedges, and node attributes), Processing (network transformations including symmetrize, dichotomize, filter, and random graph generation), Analysis (proof-of-concept functions for degree centrality, density, shortest paths, connected components, and attribute summaries), and Utilities (file I/O with support for text and binary serialization, layer import/export, and helper functions).

\subsection{One-mode Layers}

One-mode layers store edges as nodelists, where each node maintains references to its outbound and inbound edges. The specific storage type depends on whether the layer is binary or valued, and directional or symmetric. For directional layers, both incoming and outgoing edges are stored by default, but storage of inbound edges can be disabled (e.g., for random walker algorithms), reducing memory usage by nearly half.

\subsection{Two-mode Layers and Pseudo-projection}

A two-mode layer stores a set of named hyperedges, each holding a collection of node IDs connected through that affiliation. A supplementary dictionary maps nodes to their hyperedge memberships, enabling fast access to each node's affiliations. This dual indexing allows:

\begin{itemize}
    \item Finding the alters of a node (as the union of co-members across all hyperedges)
    \item Checking whether two nodes are connected (by finding overlaps in their hyperedge collections)
    \item Computing edge values (as the count of shared hyperedge memberships)
\end{itemize}

Critically, both one-mode and two-mode layer classes implement a shared interface with identical method signatures for querying edges. This means two-mode layers can be queried as if they were projected into their one-mode counterparts—without ever materializing the projection. The following C\# code illustrates the core pseudo-projection methods:

\begin{lstlisting}[language={[Sharp]C},caption={Pseudo-projection methods in LayerTwoMode class}]
/// Checks if two nodes share at least one hyperedge
public bool CheckEdgeExists(uint node1Id, uint node2Id)
{
  if (GetNonEmptyHyperedgeCollection(node1Id) is not HyperedgeCollection col1
    || GetNonEmptyHyperedgeCollection(node2Id) is not HyperedgeCollection col2)
    return false;
  // O(min(n,m)) using HashSet.Overlaps(), stops at first match
  return col1.HyperEdges.Overlaps(col2.HyperEdges);
}

/// Gets the number of shared hyperedges (pseudo-projected edge value)
public float GetEdgeValue(uint node1Id, uint node2Id)
{
  if (GetNonEmptyHyperedgeCollection(node1Id) is not HyperedgeCollection col1
    || GetNonEmptyHyperedgeCollection(node2Id) is not HyperedgeCollection col2)
    return 0f;
  // O(min(n,m)) - iterates smaller set, O(1) lookup in larger
  return col1.HyperEdges.Count < col2.HyperEdges.Count ?
    col1.HyperEdges.Count(h => col2.HyperEdges.Contains(h)) :
    col2.HyperEdges.Count(h => col1.HyperEdges.Contains(h));
}

/// Gets all nodes sharing hyperedge membership (pseudo-projected alters)
public uint[] GetNodeAlters(uint nodeId, EdgeTraversal edgeTraversal)
{
  if (GetNonEmptyHyperedgeCollection(nodeId) is not HyperedgeCollection col)
    return [];
  HashSet<uint> alters = [];
  foreach (Hyperedge hyperEdge in col.HyperEdges)
    alters.UnionWith(hyperEdge.NodeIds);
  alters.Remove(nodeId);
  return [.. alters];
}
\end{lstlisting}

\subsection{Threadle.CLIconsole and threadleR}

Threadle.CLIconsole is a cross-platform command-line application exposing Core functionality through its dedicated scripting language. It operates in two modes: human-readable text mode for interactive use, and JSON mode enabling programmatic control from external systems.

The threadleR package (\url{https://github.com/YukunJiao/threadleR}) leverages JSON mode to provide seamless R integration, wrapping Threadle commands in R functions and enabling users to combine Threadle's efficient storage with R's statistical capabilities and conditional program flows. This architecture allows researchers to implement complex sampling and traversal-based analytical workflows while Threadle handles the memory-intensive storage layer.

\section{Performance Evaluation}

To demonstrate Threadle's memory efficiency and query performance, we constructed a synthetic benchmark network representing a realistic population-scale scenario with four relational layers, each generated using a different random graph model:

\begin{itemize}
    \item \textbf{Erdős-Rényi} \cite{erdos_renyi_1959}: Each possible edge exists independently with probability $p$, producing a uniform random baseline with Poisson degree distribution. The Threadle implementation for Erdős-Rényi networks uses the one proposed by Batagelj and Brandes \cite{batagelj_efficient_2005}.
    
    \item \textbf{Watts-Strogatz} \cite{watts_strogatz_1998}: Starts from a regular ring lattice where each node connects to $k$ neighbors, then rewires edges with probability $\beta$, producing small-world networks with high clustering and short path lengths.
    
    \item \textbf{Barabási-Albert} \cite{barabasi_albert_1999}: Nodes attach preferentially to well-connected nodes, producing scale-free networks with power-law degree distributions characteristic of many real-world networks.
    
    \item \textbf{Random two-mode}: Generates $h$ hyperedges and assigns each node to affiliations drawn from a Poisson distribution with mean $a$.
\end{itemize}

The script in Listing~\ref{lst:benchmark} generates the complete benchmark network. Note that this requires approximately 32 GB of available RAM and may take a couple of minutes to complete.

\begin{lstlisting}[numbers=left,numberstyle=\tiny\color{gray},caption={Benchmark network generation script},label=lst:benchmark]
# Create nodeset with 20 million nodes
nodes = createnodeset(createnodes = 20000000)
net = createnetwork(nodeset = nodes)

# Layer 1: Erdos-Renyi random graph (~200M edges)
addlayer(net, "Random", mode = 1, directed = false)
generate(net, "Random", type = er, p = 0.000001)

# Layer 2: Watts-Strogatz small-world (200M edges)
addlayer(net, "Neighbors", mode = 1, directed = false)
generate(net, "Neighbors", type = ws, k = 20, beta = 0.1)

# Layer 3: Barabasi-Albert scale-free (~200M edges)
addlayer(net, "Communication", mode = 1, directed = false)
generate(net, "Communication", type = ba, m = 10)

# Layer 4: Random two-mode affiliations
addlayer(net, "Workplaces", mode = 2)
generate(net, "Workplaces", type = 2mode, h = 10000, a = 20)

# Save in compressed binary format (optional; can take a while)
savefile(nodes, file = "benchmark_nodes.bin.gz")
savefile(net, file = "benchmark_net.bin.gz")
\end{lstlisting}

After generating or loading the network, readers can verify query performance with commands such as:

\begin{lstlisting}[numbers=left,numberstyle=\tiny\color{gray},caption={Example query commands}]
# Check if two nodes share an affiliation (pseudo-projected edge)
checkedge(net, Workplaces, 1000000, 5000000)

# Get number of shared affiliations (pseudo-projected edge value)
getedge(net, Workplaces, 1000000, 5000000)

# Get alters in a specific layer
getnodealters(net, 1000000, layernames = Workplaces)

# Get alters across multiple layers
getnodealters(net, 1000000, layernames = Random;Neighbors;Communication)

# Get alters across multiple layers of different modes
getnodealters(net, 1000000, layernames = Workplaces;Communication)

# Get alters across all layers
getnodealters(net, 1000000)

# Shortest path across all layers (including pseudo-projected two-mode)
shortestpath(net, 1000000, 5000000)

# Shortest path in a single one-mode layer (200 million edges):
shortestpath(net, 1000000, 5000000, layername = Neighbors)

\end{lstlisting}

Edge existence checks, edge value queries and alter retrievals are effectively instantaneous, for both one-mode and two-mode layers. Shortest path calculations, which must traverse the network, range from sub-second to several seconds, depending on the number and types of layers that are to be explored. Table~\ref{tab:benchmark} summarizes the configuration and its memory footprint.

\begin{table}[ht]
\centering
\caption{Benchmark configuration and performance metrics}
\label{tab:benchmark}
\begin{tabular}{@{}lr@{}}
\toprule
\textbf{Configuration} & \textbf{Value} \\
\midrule
Total nodes & 20,000,000 \\
Layer 1: Erdős-Rényi (symmetric) & $\sim$200,000,000 edges \\
Layer 2: Watts-Strogatz (symmetric) & 200,000,000 edges \\
Layer 3: Barabási-Albert (symmetric) & $\sim$200,000,000 edges \\
Layer 4: Two-mode hyperedges & 10,000 \\
Layer 4: Affiliations per node & 20 \\
Mean nodes per hyperedge & 40,000 \\
Equivalent projected edges & $\sim$8 trillion \\
\midrule
\textbf{Memory} & \\
\midrule
RAM usage (Windows 11, x64) & 20 GB \\
Disk storage (compressed binary) & 2.9 GB \\
\bottomrule
\end{tabular}
\end{table}

\subsection{Memory Analysis}

The two-mode layer with 10,000 hyperedges and 20 affiliations per node across 20 million nodes contains 400 million node-affiliation ties, averaging 40,000 nodes per hyperedge. Projecting this to one-mode form would produce approximately:

\begin{equation}
10{,}000 \times \frac{40{,}000 \times 39{,}999}{2} \approx 8 \times 10^{12} \text{ edges}
\end{equation}

At 8 bytes per edge (two 32-bit node IDs), materializing this projection would require approximately 64 terabytes of memory—clearly infeasible.

Threadle stores this same relational information in approximately 20 GB total (including the 600 million one-mode edges across three layers as well as the client itself), representing a compression ratio exceeding \textbf{2000:1} for the two-mode layer alone. This dramatic reduction is achieved because the pseudo-projection approach stores only the hyperedge memberships (400 million node-affiliation ties) rather than all pairwise relationships.

\subsection{Query Performance}

Despite storing 2-mode data as hyperedges, query operations remain fast. Checking edge existence between two nodes, retrieving edge values, and fetching all alters are seemingly instantaneous. This performance is achieved through:

\begin{itemize}
    \item Hash-based lookups for hyperedge membership (O(1) average case)
    \item Set intersection optimizations that iterate the smaller collection
    \item Early termination for existence checks (stops at first shared hyperedge)
\end{itemize}

For sample- and traversal-based methods—random walks, ego network extraction, neighborhood sampling—this query performance is sufficient to support interactive analysis rates.

\section{Research Application}

Threadle was developed within \textit{The Complete Network of Sweden} (NetReg, \url{https://netreg.se}) research environment, which aims to construct and analyze the comprehensive, full-population network of social exposure in Sweden using administrative register data. This network encompasses kinships, households, families, and affiliations (schools, workplaces, residential areas) for approximately 15 million individuals from 1990 onward, representing one of the most complete, multi-domain population-scale social network datasets available for research.

Using Threadle, we are developing and testing various random-walker methodologies for measuring network properties across multilayer population networks. These methods, implemented in threadleR, enable researchers to estimate network statistics through sampling and traversal rather than exhaustive computation—essential for networks where exhaustive computation is infeasible.

Beyond administrative data, Threadle's design is applicable to any domain requiring efficient storage and querying of large, multilayer, mixed-mode networks: biological interaction networks, infrastructure systems, bibliometric networks, or any context where two-mode affiliations must be queried at scale without materialization.

\section{Limitations and Future Work}

Threadle is designed primarily as a storage and query engine, not a comprehensive analytical toolkit. While it includes some basic analytical functions as proof of concept—degree centrality, density, connected components, and shortest path calculations—more advanced metrics (clustering coefficients, betweenness centrality, community detection) that are well-served by existing libraries and often computationally infeasible at population scale are not implemented.

Current limitations include:

\begin{itemize}
    \item \textbf{Single-machine architecture}: Threadle operates in-memory on a single machine. Distributed or out-of-core processing is not currently supported.
    
    \item \textbf{Static networks}: The current implementation is optimized for networks that are loaded, queried, and potentially modified, but not for streaming or highly dynamic edge updates.
    
    \item \textbf{Focus on storage over analytics}: While basic metrics are available, complex analytical methods are intended to be implemented at the frontend level (e.g., in threadleR) or by exporting data to specialized tools.
\end{itemize}

Planned extensions include:

\begin{itemize}
    \item Additional analytical functions in threadleR focusing on multilayer and attribute-rich methods
    \item Sample- and traversal-based algorithms for network statistics estimation
    \item Enhanced support for temporal network sequences
\end{itemize}

\section{Availability and Documentation}

Threadle is open-source software released under the MIT license. Precompiled binaries are available for Windows, Linux, and macOS, and the software can be compiled from source.

\begin{itemize}
    \item Project website: \url{https://www.threadle.dev}
    \item Source code: \url{https://github.com/carlnordlund/Threadle}
    \item threadleR: \url{https://github.com/YukunJiao/threadleR}
\end{itemize}

The project website provides comprehensive documentation including installation instructions, a quick-start guide, technical architecture details, and a complete reference to all CLI commands. The user guide covers data structures (Nodesets, Networks, layers), file I/O operations with multiple format options (.tsv, .tsv.gz, .bin, .bin.gz), and configurable settings for memory optimization. A vignette using the Lazega lawyer network data \cite{lazega_collegial_2001} demonstrates threadleR usage.

\section{Acknowledgements}

Threadle was developed within the research environment The Complete Network of Sweden (NetReg), funded by the Swedish Research Council (grant 2024-01861).

\bibliographystyle{plain}

\end{document}